# Magnetic Order versus superconductivity in the Iron-based layered $La(O_{1-x}F_x)FeAs$ systems

Clarina de la Cruz [1,2], Q. Huang[3], J. W. Lynn[3], Jiying Li[3,4], W. Ratcliff II[3], J. L. Zarestky[5], H. A. Mook[2], G. F. Chen[6], J. L. Luo[6], N. L. Wang[6], and Pengcheng Dai[1,2]

[1] Department of Physics and Astronomy, The University of Tennessee, Knoxville, Tennessee 37996-1200, USA

[2]Oak Ridge National Laboratory, Oak Ridge, Tennessee 37831, USA

[3] NIST Center for Neutron Research, National Institute of Standards and Technology, Gaithersburg, Maryland 20899-8562 USA

[4] Department of Materials Science and Engineering, University of Maryland, College Park, Maryland 20742-6393 USA

[5]Ames Laboratory and Department of Physics and Astronomy, Iowa State University, Ames, Iowa 50011, USA

[6] Beijing National Laboratory for Condensed Matter Physics, Institute of Physics, Chinese Academy of Sciences, Beijing 100080, Peoples Republic of China

**In high-transition temperature (high-$T_c$) copper oxides, it is generally believed that antiferromagnetism plays a fundamental role in the superconducting mechanism because superconductivity occurs when mobile 'electrons' or 'holes' are doped into the antiferromagnetic parent compounds[1,2]. The recent discovery of superconductivity in the rare-earth (*R*) iron-based oxide systems [$RO_{1-x}F_xFeAs$ (refs. 3-7)] has generated enormous interest because these materials are the first non-** **copper oxide superconductors with $T_c$ exceeding 50 K. The parent (non-superconducting) LaOFeAs material is metallic but shows anomalies near 150 K in both resistivity and dc magnetic susceptibility[3]. While optical conductivity and theoretical calculations suggest that LaOFeAs exhibits a spin-density-wave (SDW) instability that is suppressed with doping electrons to form superconductivity[8], there has been no direct evidence of the SDW order. Here we use neutron scattering to demonstrate that LaOFeAs undergoes an abrupt structural distortion below ~150 K, changing the symmetry from tetragonal (space group *P4/nmm*) to monoclinic (space group *P112/n*) at low temperatures, and then followed with the development of long range SDW-type antiferromagnetic order at ~134 K with a small moment but simple magnetic structure[8]. Doping the system with flourine suppresses both the magnetic order and structural distortion in favor of superconductivity. Therefore, much like high-$T_c$ copper oxides, the superconducting regime in these Fe-based materials occurs in close proximity to a long-range ordered antiferromagnetic ground state.**

Since the discovery of long-range antiferromagnetic order in the parent compounds of high-$T_c$ superconductors,[1,2] there have been tremendous efforts in understanding the relation of magnetism to the superconductivity of these materials. Much like high-$T_c$ copper oxides, superconductivity in the newly discovered Fe-based materials are derived from either electron[3,5,6,7] or hole[4] doping of their nonsuperconducting parent compounds. The question immediately comes to mind, then, is what is the ground state of the parent compounds? Theoretically, it has been argued that nonsuperconducting LaOFeAs is either a nonmagnetic metal near a magnetic (antiferromagnetic and/or ferromagnetic) instability[9,10,11] or an antiferromagnetic semimetal[8,12, 13]. As a function of temperature, the resistivity of LaOFeAs shows a clear drop around 150 K before turning back up below 50 K (refs. 3, 8). The DC magnetic susceptibility also has a small anomaly near 150 K. From optical measurements and theoretical calculations[8], it is argued that LaOFeAs has an antiferromagnetic SDW instability below 150 K and superconductivity in these materials arises from the suppression of the SDW order.

We used neutron diffraction to study the structural and magnetic order in the polycrystalline nonsuperconducting LaOFeAs and superconducting $La(O_{1-x}F_x)FeAs$ with $x = 0.08$ ($T_c = 26$ K). We prepared ~2 grams each of these samples using the method described in Ref. 8. Our experiments were carried out on the BT-1 powder diffractometer and BT-7 thermal triple-axis spectrometer at the NIST Center for Neutron Research, Gaithersburg, Maryland, and on the HB-1A triple-axis spectrometer at the High Flux Isotope Reactor, Oak Ridge National Laboratory.

We first discuss the crystallography of these samples. Figure 1a shows the high resolution neutron powder diffraction data and our refinements for the nonsuperconducting LaOFeAs at 170 K. Consistent with earlier results[3-8], we find that the crystal structure belongs to the tetragonal *P4/nmm* space group with atomic positions given in Table 1. Upon cooling the sample to 4 K, the (2,2,0) reflection that has a single peak at 170 K (see inset in Fig. 1a) is split into two peaks (inset in Fig. 1b). This immediately suggests that a structural phase transition has occurred. For comparison, we note that the (2,2,0) peak remains a single peak even at 10 K (see inset in Figs. 1c) for superconducting $La(O_{0.92}F_{0.08})FeAs$. To understand the low temperature structural distortion in LaOFeAs, we carried out refinements of the neutron data and found that the structure in fact becomes monoclinic and belongs to space group *P112/n* . Table 2 summarizes the low temperature lattice parameters and atomic positions for LaOFeAs. Table 3 summarizes the lattice parameters and atomic positions for superconducting $La(O_{0.92}F_{0.08})FeAs$ at 10 K, 35 K, and 170 K. Figure 1d shows the LaOFeAs structure.

To see if the newly observed structural transition is related to the ~150 K resistivity anomaly, we carried out detailed temperature dependent measurement of the (2,2,0) reflection and found that there is an abrupt splitting of the (2,2,0) peak at 155 K (Fig. 2). The peak intensity also shows a clear kink at 155 K. These results thus indicate that the nonsuperconducting system has a structural phase transition and this phase transition is associated with the observed resistivity[3] and specific heat anomalies[8]. Since a similar splitting of the (2,2,0) peak is absent in superconducting $La(O_{0.92}F_{0.08})FeAs$ (Fig. 1c), one can safely assume that this transition is suppressed with the appearance of superconductivity in $La(O_{1-x}F_x)FeAs$ via F-doping.

If resistivity[3], specific heat[8], and neutron scattering data (Fig. 2) all indicate a phase transition near 150 K for the nonsuperconducting LaOFeAs, it would be interesting to see whether this phase transition is indeed associated with SDW order. Figure 3a shows our raw data for LaOFeAs collected on the BT-7 triple axis spectrometer at 8 K and 170 K. Inspection of the figure immediately reveals that there are extra peaks in the low temperature spectrum at wavevectors Q = 1.15, 1.53, and 2.5 $Å^{-1}$ that are not present at 170 K. Indexing these peaks in Fig. 3a indicates that these reflections are indeed directly related to the nuclear structure and are magnetic scattering arising from a simple stripe-type antiferromagnetic structure of Fe moments with a magnetic cell $\sqrt{2}a_N \times \sqrt{2}b_N \times 2c_N$ (see upper right inset of Fig. 4). Figure 3c shows our refinements considering both the magnetic and structural unit cell. Normalizing the magnetic intensity to the nuclear scattering, we find an ordered Fe moment of 0.36(5) $\mu_B$ at 8 K. For comparison, Figure 3b plots the HB-1A data for LaOFeAs and $La(O_{0.92}F_{0.08})FeAs$. For nonsuperconducting LaOFeAs, the temperature difference spectrum (between 8 K and 170 K) shows a clear peak at Q = 1.53 $Å^{-1}$ which corresponds to the magnetic (1,0,3) Bragg peak. The identical scan in the superconducting $La(O_{0.92}F_{0.08})FeAs$ shows that this peak is absent.

To see if the observed magnetic scattering at low temperature in LaOFeAs is indeed associated with the 150 K phase transition, we carried out order parameter measurement on the strongest (1,0,3) magnetic peak on both BT-7 and HB-1A. Figure 4 shows the temperature dependence of the ordered magnetic moment squared (normalized at low temperature) which vanishes at 134 K, about ~21 K lower than the structural phase transition (Fig. 2). Surprisingly, the magnetic order is established at lower temperatures than the structural distortion, much like spin ordering is established after the charge ordering in the static stripe-ordered copper oxide material $La_{1.6-x}Nd_{0.4}Sr_xCuO_4$ with $x$ = 0.12 (ref. 14). The presence of the lattice distortion above the Neel temperature is established conclusively in the lower left inset of Figure 4, where one sees clear lattice distortion at 138 K. The upper inset in Figure 4 shows our determined magnetic structure.

To summarize, we have discovered that the parent compound of the Fe-based superconductors is a long-range ordered antiferromagnet with a simple stripe-type antiferromagnetic structure within the plane that is doubled along the $c$-axis (upper right inset in Fig. 4 ). There is a structural phase transition before the antiferromagnetic phase transition that changes the structure from space group *P4/nmm* to *P112/n* at low temperature. The magnetic structure is consistent with theoretical prediction[8], but the 0.36(5) $\mu_B$ per Fe moment we observe at 8 K is much smaller than the predicted value of ~2.3 $\mu_B$ per Fe (refs. 11,12). The disappearance of the static antiferromagnetic order and lattice distortion in the doped superconducting materials suggests that the underlying physical properties of this class of superconductors may have important similarities to the high-$T_c$ copper oxides. Nevertheless, there is no doubt that this new class of materials will open new avenues of research regardless of the origin for the electron pairing and superconductivity.

**Acknowledgements** We thank J. A. Fernandez-Baca and C. Brown for helpful discussions. This work is supported by the US DOE Division of Materials Science, Basic Energy Sciences. This work is also supported by the US DOE through UT/Battelle LLC.

**Author Information** Reprints and permissions information is available at npg.nature.com/reprintsandpermissions. The authors declare no competing financial interests. Correspondence and requests for materials should be addressed to P.D. (daip@ornl.gov).

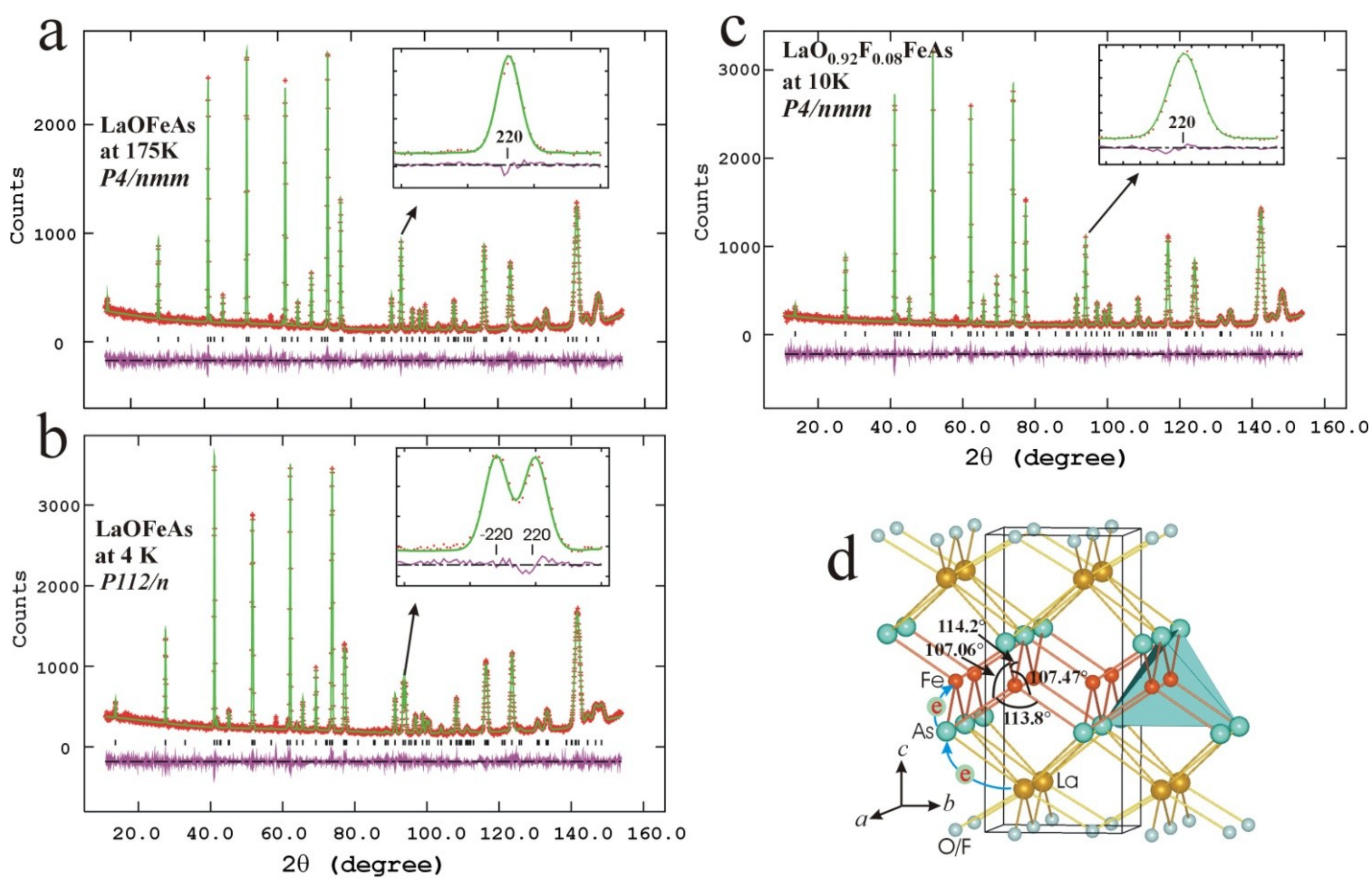

**Figure 1 Temperature dependence of the nuclear structures for LaOFeAs and $La(O_{0.92}F_{0.08})FeAs$.** The data were collected on the BT-1 diffractometer with Ge(3,1,1) monochromator and an incident beam wavelength $\lambda = 2.0785$ Å. a) Observed (crosses) and calculated (solid line) neutron powder diffraction intensities of LaOFeAs at 175 K using space group *P4/nmm*. The inset shows a single peak of the (2,2,0) reflection. Vertical lines show the Bragg peak positions. b) The same scan at 4 K, where the (2,2,0) reflection is split due to the monoclinic distortion. The fit is using space group *P112/n*. c) The 10 K scan on superconducting $La(O_{0.92}F_{0.08})FeAs$, where the space group *P4/nmm* can describe data at all measured temperatures. d) Crystal structure of $La(O_{1-x}F_x)FeAs$. For $x = 0$, the compound has the charge balance configuration $La^{3+}O^{2-}Fe^{2+}As^{3-}$. Electron doping can be achieved by replacing O with F.

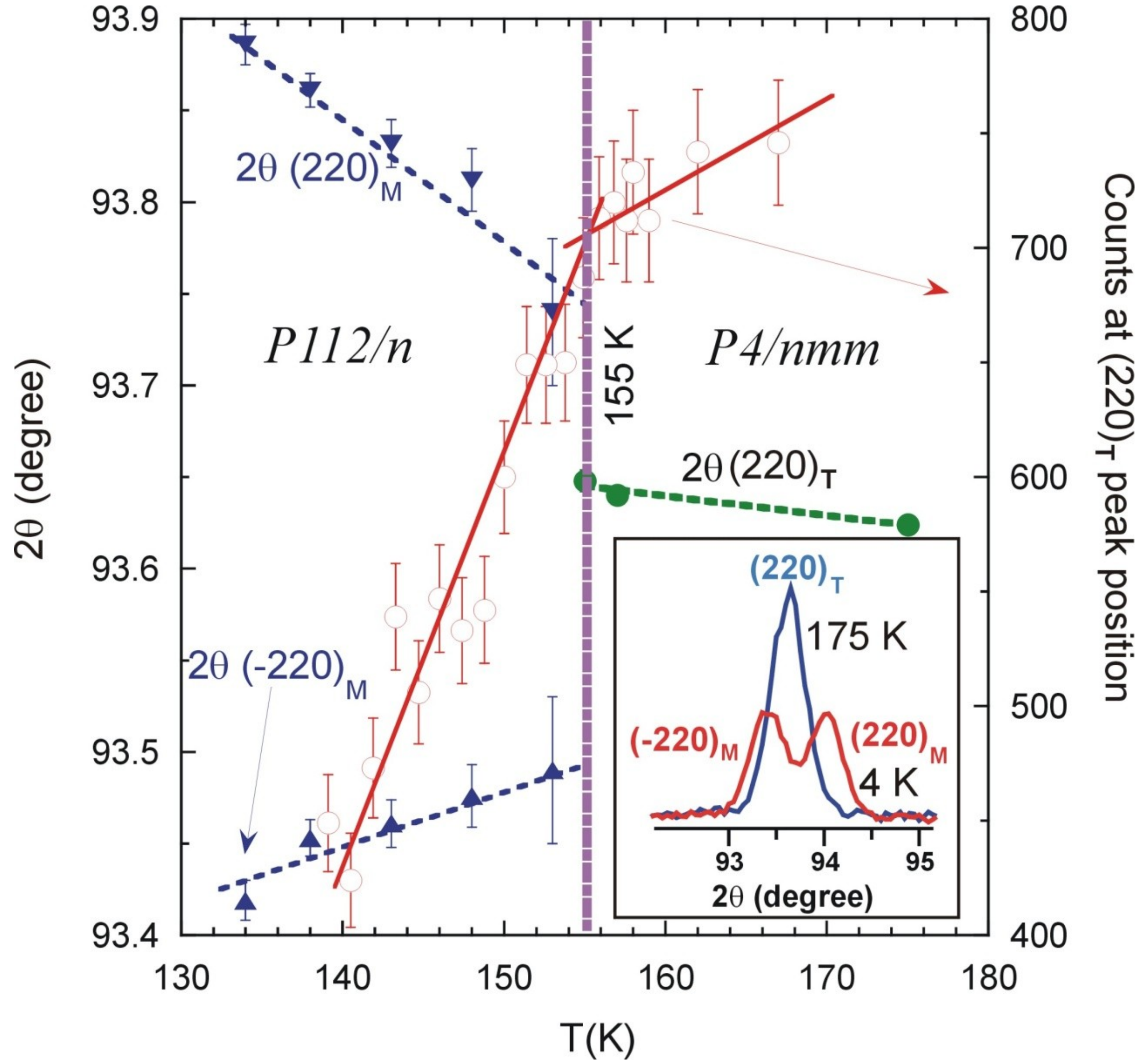

**Figure 2 Temperature dependence of the (2,2,0) nuclear reflection indicative of a structural phase transition at ~155 K in LaOFeAs.** Peak intensities at the $(2,2,0)_T$ (T denotes tetragonal) reflection (open symbol) and peak position of $(2,2,0)_T$, $(-2,2,0)_M$ (M denotes monoclinic), and $(2,2,0)_M$ (solid symbols) as a function of temperature on cooling (open symbols). A structural transition from tetragonal symmetry *P4/nmm* to monoclinic symmetry *P112/n* occurs at ~155 K. Inset shows the $(2,2,0)_T$ reflection at 175 K, and $(-2,2,0)_M$ and $(2,2,0)_M$ reflections at 4K.

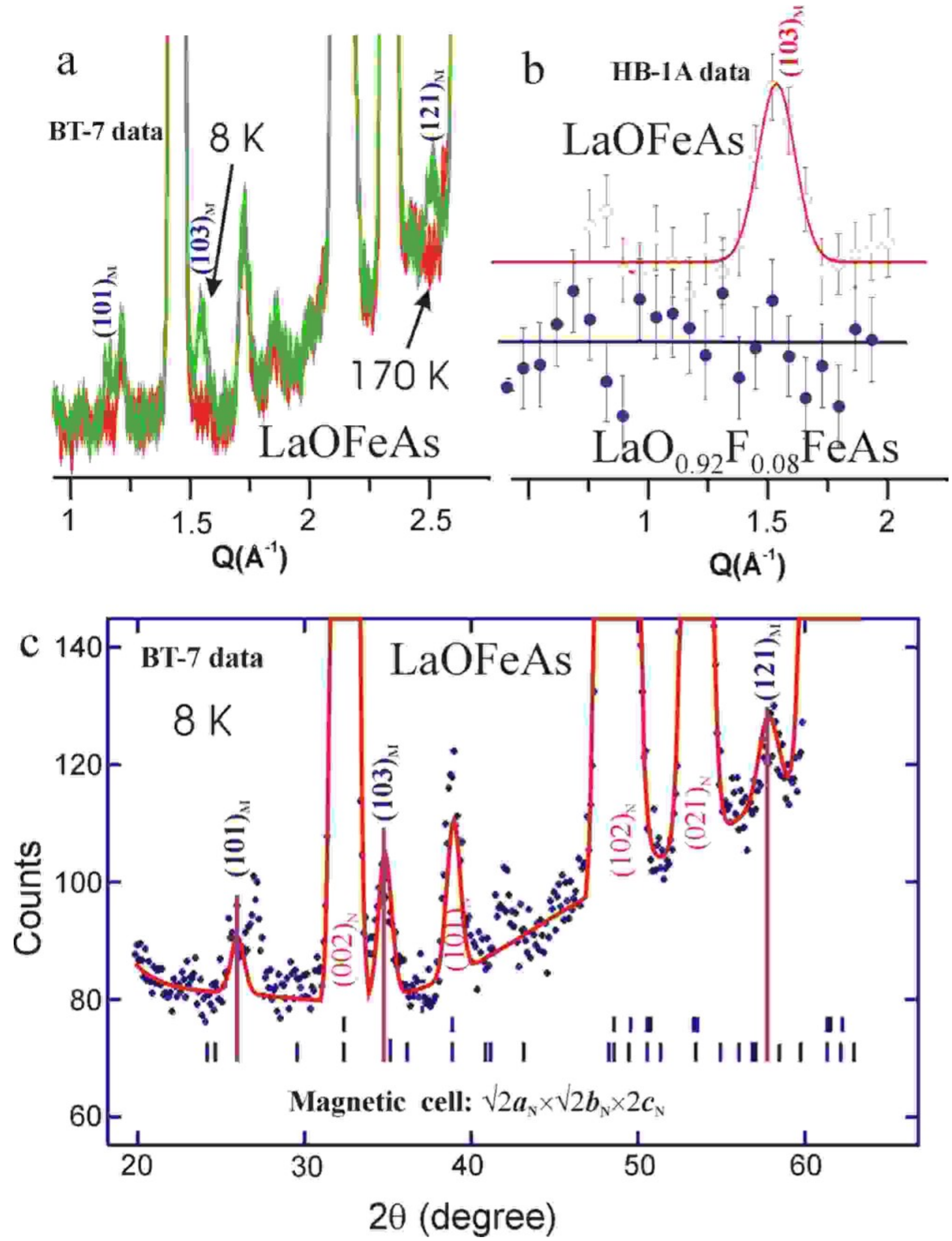

**Figure 3 Temperature dependence of the magnetic scattering for LaOFeAs and $La(O_{0.92}F_{0.08})FeAs$.** Data in panels a) and c) were collected on BT-7 with an incident beam wavelength $\lambda = 2.44$ Å with PG(002) as monochromator and PG filter. Data in b) were collected on HB-1A with $\lambda = 2.36$ Å and PG filter. a) One can see clear (marked) magnetic peaks at 8 K that disappear at 170 K, counting 1 minute/point b) The temperature difference spectra (8 K – 170 K) taken on HB-1A for LaOFeAs and $La(O_{0.92}F_{0.08})FeAs$, counting 4 minutes/point. The magnetic (1,0,3) peak is missing in $La(O_{0.92}F_{0.08})FeAs$. c) BT-7 data again showing both magnetic and nuclear Bragg peaks together with the model fit.

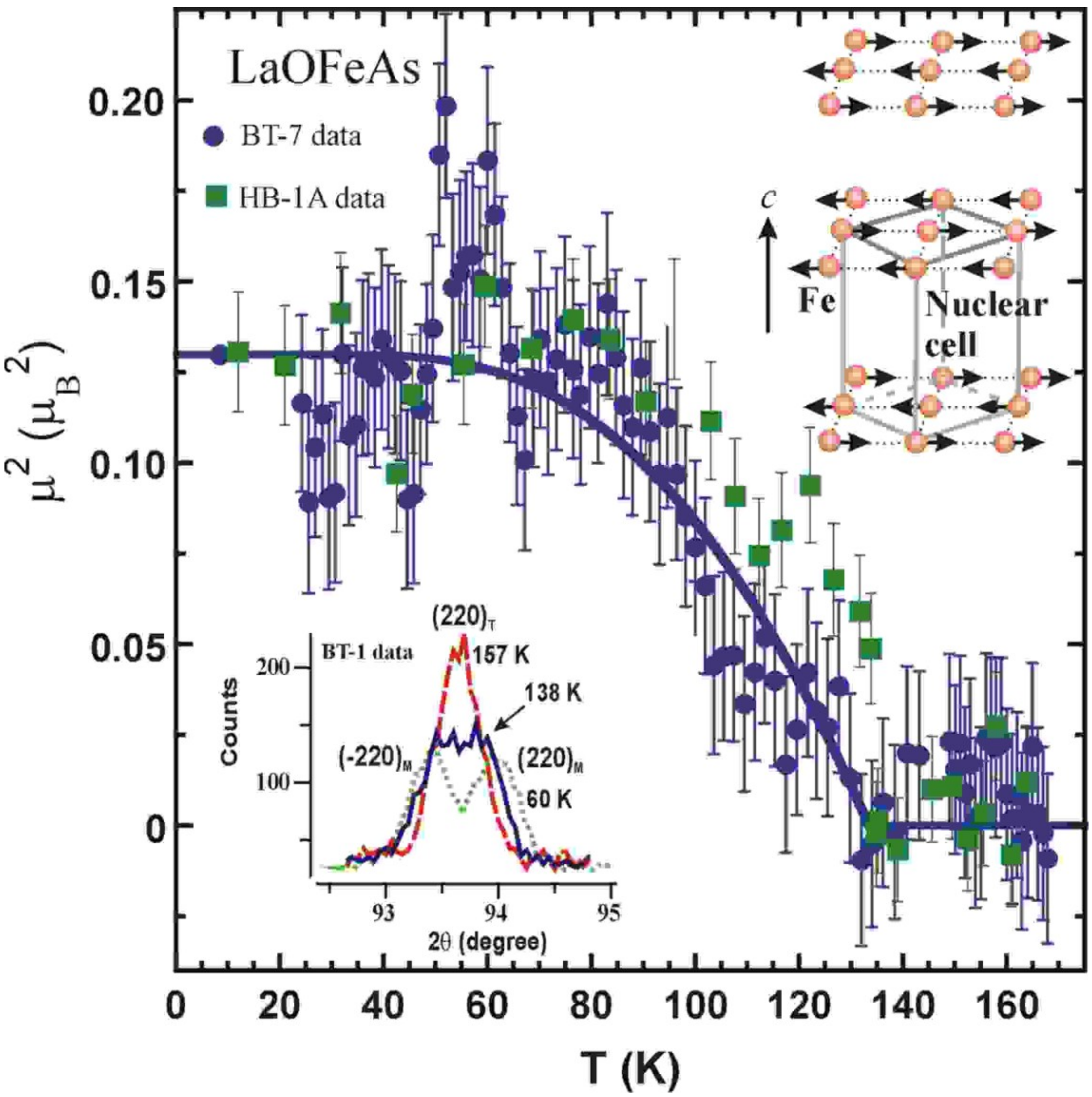

**Figure 4 Temperature dependence of the order parameter at Q = 1.53 Å$^{-1}$ obtained on BT-7/HB-1A and our determined magnetic structure for LaOFeAs.** The solid blue circles are data obtained on BT-7 and solid green squares are data taken on HB-1A. The solid line is a simple fit to mean field theory which gives $T_N$ = 134(1) K. The lower left inset shows the temperature dependence of the nuclear (2,2,0) peak obtained on BT-1. It is clear that the lattice is distorted at 138 K, before the long-range static antiferromagnetic order sets in at $T_N$ = 134(1) K. The upper right inset shows the antiferromagnetic structure of the system, giving a $\sqrt{2}a_N \times \sqrt{2}b_N \times 2c_N$ unit cell with moment directions that lie within the Fe-plane.

Table 1. Refined structure parameters of LaOFeAs at 175 K. Space group: *P4/nmm*. *a*=4.03007 (9) Å, *c*=8.7368(2) Å, *V*=141.898(9) Å$^3$.

| Atom | site | *x* | *y* | *z* | *B*(Å$^2$) |
|---|---|---|---|---|---|
| La | 2*c* | ¼ | ¼ | 0.1418(3) | 0.65(7) |
| Fe | 2*b* | ¾ | ¼ | ½ | 0.39(5) |
| As | 2*c* | ¼ | ¼ | 0.6507(4) | 0.23(8) |
| O | 2*a* | ¾ | ¼ | 0 | 0.69(7) |

Selected interatomic distances (Å) and angles (°).

| | | | |
|---|---|---|---|
| La-As ×4 | 3.378(1) | Fe-As ×4 | 2.407(2) |
| La-O ×4 | 2.365(2) | As-Fe-As | 107.41(7) |
| Fe-Fe | 2.84969(7) | As-Fe-As | 113.7(1) |

*Rp*=5.24%, *wRp*=6.62%, $\chi^2$=0.9821.

Table 2. Refined structure parameters of LaOFeAs at 4 K. Space group: *P112/n*. *a*=4.0275 (2) Å, *b*=4.0275(2) Å, *c*=8.7262(5) Å, γ=90.279(3) °, *V*=141.54(2) Å$^3$.

Lattice constants *a* and *b* were constrained to be equal in the final refinement.

| Atom | site | *x* | *y* | *z* | *B*(Å$^2$) |
|---|---|---|---|---|---|
| La | 2*e* | ¼ | ¼ | 0.1426(3) | 0.54(6) |
| Fe | 2*f* | ¾ | ¼ | 0.5006(12) | 0.16(4) |
| As | 2*e* | ¼ | ¼ | 0.6499(4) | 0.23(7) |
| O | 2*f* | ¾ | ¼ | -0.0057(17) | 0.69(7) |

Selected interatomic distances (Å) and angles (°).

| | | | |
|---|---|---|---|
| La-As ×2 | 3.369(1) | Fe-As ×2 | 2.398(6) |
| La-As ×2 | 3.380(1) | Fe-As ×2 | 2.405(6) |
| La-O ×2 | 2.394(8) | As-Fe-As | 114.2(5) |
| La-O ×2 | 2.342(7) | As-Fe-As | 107.47(6) |
| Fe-Fe | 2.8409(2) | As-Fe-As | 107.06(6) |
| Fe-Fe | 2.8548(2) | As-Fe-As | 113.8(4) |

*Rp*=4.31%, *wRp*=5.74%, $\chi^2$=1.100.

Table 3. Refined structure parameters of $LaO_{0.92}F_{0.08}FeAs$ at 10 (first line), 35 (second line), and 175 (third line) K. Space group: *P4/nmm*.

$a$=4.02016(8) Å, $c$=8.7034(2) Å, $V$=140.662(8) Å$^3$.

| | | |
|---|---|---|
| 4.01956(8) | 8.7027(2) | 140.609(8) |
| 4.02288(8) | 8.7142 | 141.027(8) |

| Atom | site | *x* | *y* | *z* | *B*(Å$^2$) |
|---|---|---|---|---|---|
| La | 2*c* | ¼ | ¼ | 0.1448(3) | 0.40(5) |
| | | ,, | ,, | 0.1458(3) | 0.50(5) |
| | | ,, | ,, | 0.1446(3) | 0.73(5) |
| Fe | 2*b* | ¾ | ¼ | ½ | 0.32(4) |
| | | ,, | ,, | ,, | 0.41(4) |
| | | ,, | ,, | ,, | 0.65(4) |
| As | 2*c* | ¼ | ¼ | 0.6521(4) | 0.41(7) |
| | | ,, | ,, | 0.6515(4) | 0.40(6) |
| | | ,, | ,, | 0.6527(4) | 0.69(7) |
| O/F | 2*a* | ¾ | ¼ | 0 | 0.53(6) |
| | | ,, | ,, | ,, | 0.62(6) |
| | | ,, | ,, | ,, | 0.71(6) |

Selected interatomic distances (Å) and angles (°).

| | | | |
|---|---|---|---|
| La-As ×4 | 3.347(1) | Fe-As ×4 | 2.407(2) |
| | 3.345(1) | | 2.404(2) |
| | 3.349(1) | | 2.412(2) |
| La-O/F×4 | 2.373(2) | As-Fe-As | 107.61(6) |
| | 2.377(1) | | 107.52(6) |
| | 2.373(1) | | 107.72(6) |
| Fe-Fe | 2.84268(6) | As-Fe-As | 113.3(1) |
| | 2.84226(6) | | 113.5(1) |
| | 2.84460(6) | | 113.0(1) |

*Rp*=5.34%, *wRp*=6.95%, $\chi^2$=1.028;

| | | |
|---|---|---|
| 5.38% | 6.96% | 1.050; |
| 5.30% | 6.93% | 0.9882. |